\documentclass[english,aps,reprint,twocolumn,superscriptaddress,longbibliography]{revtex4-2}
\usepackage[english]{babel}

\usepackage{amsmath,amssymb}
\usepackage{graphicx,xcolor}
\usepackage[unicode,bookmarksnumbered,breaklinks,colorlinks,linktocpage,
citecolor=blue,linkcolor=darkred,urlcolor=darkblue]{hyperref}
\usepackage{bm}
\usepackage{hyperref}

\definecolor{darkblue}{rgb}{0.0, 0.0, 0.55}
\definecolor{darkgreen}{rgb}{0.0, 0.2, 0.13}
\definecolor{darkred}{rgb}{0.55, 0.0, 0.0}

\begin{document}

\title{Quantum Heisenberg antiferromagnet in a field on the Tasaki square lattice}

\author{Maksym Parymuda}
\affiliation{Yukhnovskii Institute for Condensed Matter Physics 
	of the National Academy of Sciences of Ukraine,\\
    Svientsitskii Street 1, 79011 L'viv, Ukraine}

\author{Taras Krokhmalskii} 
\affiliation{Yukhnovskii Institute for Condensed Matter Physics 
	of the National Academy of Sciences of Ukraine,\\
	Svientsitskii Street 1, 79011 L'viv, Ukraine}

\author{Oleg Derzhko}
\affiliation{Yukhnovskii Institute for Condensed Matter Physics 
	of the National Academy of Sciences of Ukraine,\\
	Svientsitskii Street 1, 79011 L'viv, Ukraine}
\affiliation{Professor Ivan Vakarchuk Department for Theoretical Physics,
	Ivan Franko National University of L’viv,\\
	Drahomanov Street 12, 79005 L’viv, Ukraine}

\date{\today}

\begin{abstract}
We consider the $S=1/2$ Heisenberg antiferromagnet on the Tasaki square lattice (flat-band spin system) and study its low-temperature thermodynamics around the saturation magnetic field. To this end, we construct a mapping of the ground states in the subspaces with total  $S^z=N/2,\ldots,N/3$ ($N$ is the number of lattice sites) on the hard squares on an auxiliary square lattice and use classical Monte Carlo simulations to examine the latter classical system. The most prominent feature of the $S=1/2$ Heisenberg antiferromagnet on the Tasaki square lattice is an order-disorder phase transition which occurs at a low temperature just below the saturation magnetic field  and belongs to the 2D Ising universality class.
\end{abstract}


 
\maketitle

The authors dedicate this manuscript to the loving memory of Johannes Richter. The topic of localized magnons and flat-band systems was particularly close to his heart.

\section{Introduction}
\label{s1}

Frustrated quantum spin lattice systems are a subject of intense ongoing research in the field of magnetism \cite{highmagneticfields2002,quantummagnetism2004,frustratedspinsystems2005}.
Geometric frustration and quantum fluctuations act against ordering even in the  ground state and may result in exotic phases of matter. Usually, geometrical frustration makes the investigation of the quantum spin systems more difficult. In particular, the famous universal quantum Monte Carlo approach, generally speaking, fails because of the infamous sign problem. 
However, frustration may result in flat (dispersionless) magnon bands that allows for  elaboration of a completely different approach for calculation of the thermodynamic quantities. Flat-band states, also known as localized magnons, were discovered in Refs.~\cite{Schnack2001,Schulenburg2002} and their contribution to magnetothermodynamics was examined in Refs.~\cite{Zhitomirsky2004,Derzhko2006} (the interested reader may also consult two reviews, Refs.~\cite{Derzhko2007,Derzhko2015}).

In the present paper we study the spin $S=1/2$ antiferromagnetic Heisenberg model in a field on the so-called Tasaki square lattice, that is, a bond-decorated square lattice at the flat-band point, i.e., when the antiferromagnetic coupling along the decorated bond $J_2$ is 4 times larger than the antiferromagnetic coupling along the bond $J_1$ forming the basic square lattice, see the upper panel of Fig.~\ref{fig01} for a sketch of the lattice. Tasaki lattices in different dimensions were introduced in the context of the flat-band ferromagnetism of the single-orbital Hubbard model \cite{Tasaki1992} (see also \cite{Maksymenko2012,Maksymenko2015,Liu2019}). While in one dimension the Tasaki chain, also known as the sawtooth or delta chain, is well known in the context of spin models, the two-dimensional version, as far as we know, has not been examined from a spin-lattice perspective so far. With our study we shed some light on this issue. 

The paper is organized as follows. In Section~\ref{s2}, we introduce the model and explain a mapping onto a hard square problem. Then, in Section~\ref{s3}, we first consider small lattices to check the mapping by comparison with exact diagonalizations and then large ones to obtain the low-temperature magnetothermodynamics of the Tasaki-square-lattice $S=1/2$ Heisenberg antiferromagnet around the saturation magnetic field using classical statistical mechanics tools. Finally, we discuss and summarize our investigation in Section~\ref{s4}.

\section{Model. Flat-band point}
\label{s2}

\begin{figure}
\includegraphics[width=\columnwidth]{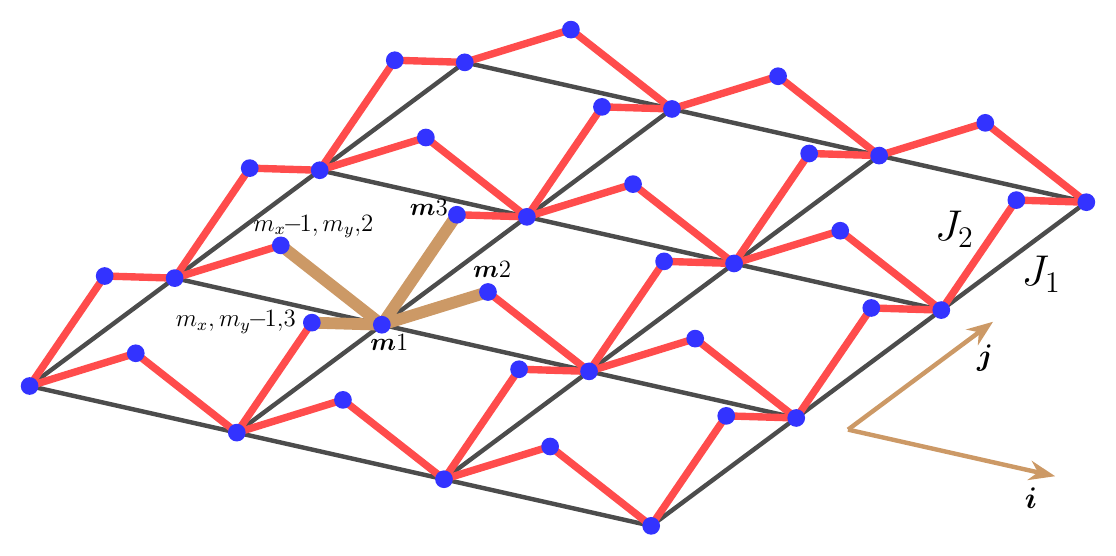}
\includegraphics[width=\columnwidth]{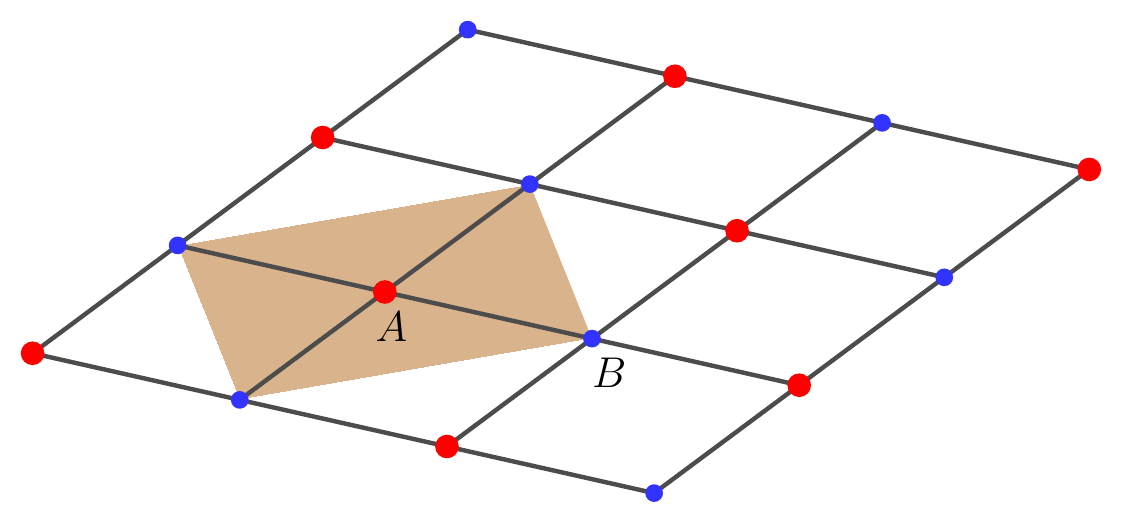}
\caption{(Top) Tasaki square lattice. Here, ${\bm m}=m_x{\bm i}+m_y{\bm j}$, $m_x$ and $m_y$ are integers, enumerates the unit cells, each of which contains three sites enumerated by $\alpha=1,2,3$. For antiferromagnetic couplings $J_2=4J_1>0$, the lowest-energy one-magnon band is dispersionless (flat). (Bottom) Auxiliary square lattice used for representation of localized magnons. It consists of two sublattices, denoted as $A$ and $B$, and all sites
of the sublattice $A$ are surrounded by the sites of the sublattice $B$ and vice versa.}
\label{fig01}
\end{figure}

The Hamiltonian of the $S=1/2$ Heisenberg lattice model under consideration reads:
\begin{equation}
\label{01}
H=\sum_{\langle\bm{m}\alpha;\bm{n}\beta\rangle}
J_{\bm{m}\alpha;\bm{n}\beta}\bm{S}_{\bm{m}\alpha}\cdot\bm{S}_{\bm{n}\beta}.
\end{equation}
The sum in Eq.~(\ref{01}) runs over the nearest bonds of the $N$-site Tasaki square lattice shown in the upper panel of Fig.~\ref{fig01}, that is, ${\bm m}=(m_x,m_y)$ runs over the unit cells (i.e., the sites of the basic square lattice) and $\alpha=1,2,3$ runs over the sites within the unit cell. The antiferromagnetic exchange coupling $J_{\bm{m}\alpha;\bm{n}\beta}$ acquires two values: (i) $J_1$ if $\alpha=\beta=1$ and $\bm{m}$ and $\bm{n}$ are the neighboring sites of the underlying square lattice or (ii) $J_2$ if the sites $\bm{m}\alpha$ and $\bm{n}\beta$ are connected by the decorated bond (colored bonds in the upper panel of Fig.~\ref{fig01}).
To obtain dispersionless (flat) one-magnon band we assume $J_1=J>0$, $J_2=4J$, see below.
It might be worth noting that the Tasaki square lattice, although may resemble in some ways the Lieb square lattice \cite{Lieb1989a,Lieb1989b}, is still different from the Lieb square lattice: The latter lattice corresponds to $J_1=0$ and $J_2=J$. 

\begin{figure}
\includegraphics[width=\columnwidth]{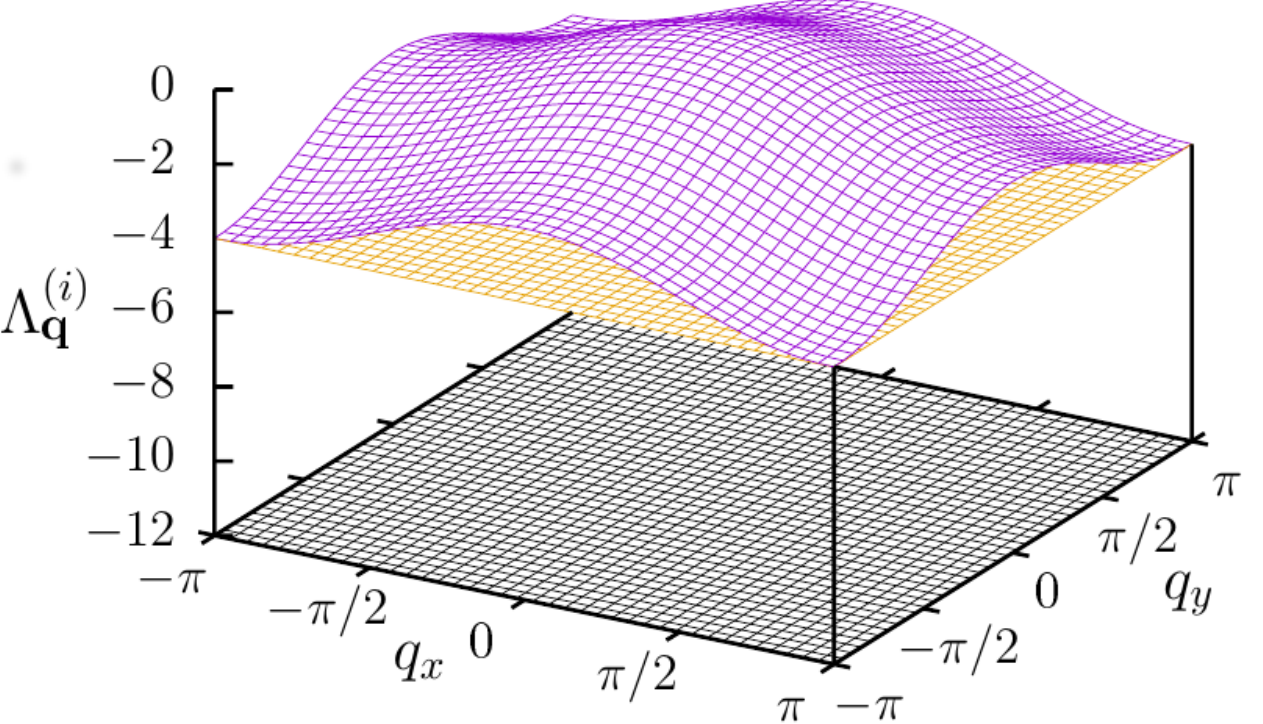}
\caption{One-magnon energies $\Lambda^{(1)}_{\bm q}$, $\Lambda^{(2)}_{\bm q}$, $\Lambda^{(3)}_{\bm q}$ given in Eq.~(\ref{04}) for $J_1=1$ and $J_2=4$.}
\label{fig02}
\end{figure}

The Hamiltonian given in Eq.~(\ref{01}) can be diagonalized in the one-magnon subspace.
To this end, we introduce
\begin{eqnarray}
\label{02}
S_{{\bm q}\alpha}^{\pm}=\frac{1}{\sqrt{{\cal N}}}\sum_{{\bm m}} e^{\mp {\rm i}{\bm q}\cdot {\bm m}} S_{{\bm m}\alpha}^{\pm},
\nonumber\\
S_{{\bm m}\alpha}^{\pm}=\frac{1}{\sqrt{{\cal N}}}\sum_{{\bm q}} e^{\pm {\rm i}{\bm q}\cdot {\bm m}} S_{{\bm q}\alpha}^{\pm},
\end{eqnarray}
where ${\bm q}=(q_x,q_y)$, $q_{x,y}=2\pi z_{x,y}/{\cal N}$, $z_{x,y}=1,\ldots,{\cal L}_{x,y}$, ${\cal L}_x{\cal L}_y={\cal N}$, ${\cal N}=N/3$ is the number of unit cells, 
and after some calculations arrive at
\begin{eqnarray}
\label{03}
H_{\rm 1m}{=}E_{\rm FM}
{+}\sum_{\bm q}
\left(
\begin{array}{ccc}
S_{{\bm q}1}^{-}\! & S_{{\bm q}2}^{-}\! & S_{{\bm q}3}^{-}
\end{array}
\right)
\!{\bf H}\!
\left(
\begin{array}{c}
S_{{\bm q}1}^{+} \\ S_{{\bm q}2}^{+} \\ S_{{\bm q}3}^{+}
\end{array}
\right),
\nonumber\\
E_{\rm FM}=\left(\frac{J_1}{2}{+}J_2\right){\cal N},
\nonumber\\
H_{11}=J_1(\cos q_x+\cos q_y) -2J_1-2J_2,
\nonumber\\
H_{12}=H^*_{21}=\frac{J_2}{2}\left(1+e^{-{\rm i}q_x}\right),
\nonumber\\
H_{13}=H^*_{31}=\frac{J_2}{2}\left(1+e^{-{\rm i}q_y}\right),
\nonumber\\
H_{22}=H_{33}=-J_2,
H_{23}=H_{32}=0.
\end{eqnarray}
Interestingly, the matrix ${\bf H}$ in Eq.~(\ref{03}) has always the eigenvalue $-J_2$ which corresponds to the eigenvector $0,-(1+e^{-{\rm i}q_y})/(1+e^{-{\rm i}q_x}),1$ (recall a flat band for the Lieb lattice). 
Diagonalizing the matrix ${\bf H}$ in Eq.~(\ref{03}) for $J_2=4J_1$ (flat-band point), we obtain the following one-magnon spectrum:
\begin{eqnarray}
\label{04}
\Lambda^{(1)}_{\bm q}=J_1\left(\cos q_x+\cos q_y-2\right),
\nonumber\\
\Lambda^{(2)}_{\bm q}=-4J_1,
\nonumber\\
\Lambda^{(3)}_{\bm q}=-12J_1=\epsilon_0.
\end{eqnarray}
The dispersion relations $\Lambda^{(i)}_{\bm q}$, $i=1,2,3$, given in Eq.~(\ref{04}) are illustrated in Fig.~\ref{fig02} for the set of exchange couplings $J_1=1$, $J_2=4$.

To obtain the eigenstates of the Hamiltonian $H_{\rm 1m}$, we perform analytical computations with the Maple software, see also Appendix~A.
Consider the obtained in such a way eigenstates
\begin{eqnarray}
\label{05}
\vert \psi^{(3)}_{\bm q}\rangle
=\left[\frac{-4}{1+e^{{\rm i}q_y}}S_{{\bm q}1}^-
+\frac{1+e^{{\rm i}q_x}}{1+e^{{\rm i}q_y}}S_{{\bm q}2}^-
+S_{{\bm q}3}^-
\right]\vert {\rm FM}\rangle
\nonumber\\
=\frac{1}{1{+}e^{{\rm i}q_y}}
\left[{-}4S_{{\bm q}1}^- 
{+} \left(1{+}e^{{\rm i}q_x}\right)S_{{\bm q}2}^-{+}\left(1{+}e^{{\rm i}q_y}\right)S_{{\bm q}3}^-\right]\vert {\rm FM}\rangle,
\nonumber\\
\vert {\rm FM}\rangle=\vert\uparrow_1\ldots\uparrow_N\rangle,
\end{eqnarray}
which correspond to $\Lambda^{(3)}_{\bm q}=\epsilon_0$.
With the help of Eq.~(\ref{02}), Eq.~(\ref{05}) can be cast into
\begin{eqnarray}
\label{06}
\vert \psi^{(3)}_{\bm q}\rangle
=
\frac{1}{\sqrt{{\cal N}}}
\sum_{{\bm m}}e^{{\rm i}{\bm q}\cdot {\bm m}}\frac{1}{1{+}e^{{\rm i}q_y}}l_{\bm m}\vert {\rm FM}\rangle,
\nonumber\\
l_{\bm m}={-}4S_{{\bm m}1}^-{+}S_{{\bm m}2}^-{+}S_{{\bm m}3}^-{+}S_{m_x-1,m_y,2}^-
{+}S_{m_x,m_y-1,3}^-.
\end{eqnarray} 
Since $\vert \psi^{(3)}_{\bm q}\rangle$ for all ${\bm q}$ correspond to the same $\Lambda^{(3)}_{\bm q}=\epsilon_0$,
it is clear now that the flat-band states can be viewed as the localized magnons $l_{\bm m}\vert{\rm FM} \rangle$. There are ${\cal N}$ such states labeled by ${\bm m}$. The localized magnon occupies five lattice sites: $m_x,m_y,1$; $m_x,m_y,2$; $m_x,m_y,3$; $m_x{-}1,m_y,2$; $m_x,m_y{-}1,3$ (see the upper panel of Fig.~\ref{fig01}).
The localized-magnon state energy is $\epsilon_0$, i.e., $Hl_{\bm m}\vert{\rm FM} \rangle=(E_{\rm FM}+\epsilon_0)l_{\bm m}\vert{\rm FM}\rangle$.

It is instructive to construct the localized-magnon states $l_{\bm m}\vert{\rm FM} \rangle$ in a different way \cite{Richter2004}. Let us for a while change notations,
$m_x,m_y,1\to 0$, $m_x,m_y,2\to 1$, $m_x,m_y,3\to 2$, $m_x-1,m_y,2\to 3$, $m_x,m_y-1,3\to 4$,
for brevity. Consider the five-site Hamiltonian
$H_{01234}=J_2{\bm S}_0\cdot\left({\bm S}_1+{\bm S}_2+{\bm S}_3+{\bm S}_4\right)$.
Let us seek for its (unnormalized) eigenstate $\vert\psi\rangle$, $H_{01234}\vert\psi\rangle=E\vert\psi\rangle$, in the form:
\begin{eqnarray}
\label{07}
\vert\psi\rangle=A\vert 0\rangle+\vert 1\rangle+\vert 2\rangle+\vert 3\rangle+\vert 4\rangle,
\end{eqnarray}
where
$\vert 0\rangle=\vert\downarrow_0\uparrow_1\uparrow_2\uparrow_3\uparrow_4\rangle$,
$\vert 1\rangle=\vert\uparrow_0\downarrow_1\uparrow_2\uparrow_3\uparrow_4\rangle$,
and so on. One can easily convince oneself that 
$H_{01234}\vert 0\rangle=J_2[(\vert 1\rangle+\vert 2\rangle+\vert 3\rangle+\vert 4\rangle)/2-\vert 0\rangle]$, 
$H_{01234}\vert 1\rangle=J_2(\vert 0\rangle+\vert 1\rangle)/2$,
and so on. If $\vert\psi\rangle$ (\ref{07}) is an eigenvalue of $H_{01234}$, the following two equations must be satisfied:
\begin{eqnarray}
\label{08}
\frac{A+1}{2}=E,
\;\;\;
-A+2=EA.
\end{eqnarray}
Therefore, according to Eq.~(\ref{08}), $A=1$ in Eq.~(\ref{07}) gives $E=J_2$, whereas $A=-4$ in Eq.~(\ref{07}) gives $E=-3J_2/2$. The latter case is interesting: If we add four more sites $m_x+1,m_y,1\to 1^\prime$, $m_x,m_y+1,1\to 2^\prime$, $m_x-1,m_y,1\to 3^\prime$, $m_x,m_y-1,1\to 4^\prime$ and eight more Heisenberg couplings as shown in the upper panel of Fig.~\ref{fig01} with the fine-tuned relation $J_2=4J_1$, the state 
\begin{eqnarray}
\label{09}
\vert\Psi\rangle=\left(-4\vert 0\rangle+\vert 1\rangle+\vert 2\rangle+\vert 3\rangle+\vert 4\rangle\right)\vert \uparrow\ldots\uparrow\rangle,
\end{eqnarray} 
where $\vert \uparrow\ldots\uparrow\rangle$ denotes the sites $1^\prime$, $2^\prime$, $3^\prime$, $4^\prime$ in the state $\vert\uparrow\rangle$,  
remains an eigenstate of the new nine-site Hamiltonian with the eigenvalue $-3J_1$. 
This is a result of destructive quantum interference, which occurs because of the  precise balance of the value $A=-4$ in Eq.~(\ref{09}) and the ratio $J_2/J_1=4$. The fully polarized state is also the eigenstate of the new nine-site Hamiltonian with the energy $9J_1$. Therefore, the energy of the state (\ref{09}) is lower than the fully-polarized-state energy by $12J_1$, cf. Eq.~(\ref{04}).
Extending further the nine-site Hamiltonian to the whole lattice, see the upper panel of Fig.~\ref{fig01}, one concludes that the state \eqref{09} with $\vert \uparrow\ldots\uparrow\rangle$ denoting now all sites except $0$, $1$, $2$, $3$, $4$ in the state $\vert\uparrow\rangle$ is the eigenstate of $H$ (\ref{01}) (with $J_2=4J_1$): $H\vert \Psi\rangle=(E_{\rm FM}+\epsilon_0)\vert \Psi\rangle$, $\epsilon_0=-12J_1$.

We note in passing that these arguments fail if $A=1$ in Eq.~(\ref{07}): After adding the mentioned above four more sites $1^\prime$, $2^\prime$, $3^\prime$, $4^\prime$ and eight more Heisenberg couplings, the state (\ref{09}) does not remain an eigenstate of the new nine-site Hamiltonian. Indeed, a spin-down escapes from the five-site trap, i.e., a nonzero amplitude of states with, e.g., $\vert\ldots\downarrow_{1^\prime}\ldots\rangle$
appears, leading, after all, to formation of the dispersive-band states $\vert \psi^{(1)}_{\bm q}\rangle$.  

Most importantly, because of the local nature of ${\cal N}$ one-magnon states $l_{\bm m}\vert{\rm FM}\rangle$ (\ref{06}), see the lower panel of Fig.~\ref{fig01}, we can construct {\em many-magnon ground states} placing localized magnons in distant traps. Therefore, the ground-state energy in the subspace with total $S^z=N/2-n$ is $E_{\rm FM}+n\epsilon_0$, whereas the ground-state degeneracy $g_{\cal N}(n)$ equals to the canonical partition function of $n$ hard squares on an ${\cal N}$-site square lattice. Here, $n=0,1,\ldots,{\cal N}/2$. 

The localized-magnon contribution to thermodynamics of the frustrated quantum spin system in the presence of an external magnetic field $h$ is given by
\begin{eqnarray}
\label{10}
Z_{\rm lm}(T,h,N)=\sum_{n=0}^{\frac{{\cal N}}{2}}g_{\cal N}(n)e^{-\frac{E_{\rm FM}-\frac{{N}}{2}h+n\epsilon_0+nh}{T}}
\nonumber\\
=e^{-\frac{E_{\rm FM}-\frac{{N}}{2}h}{T}}\Xi(z,{\cal N}),
\nonumber\\
\Xi(z,{\cal N})=\sum_{n=0}^{{\cal N}/2}z^ng_{\cal N}(n),
\;\;\;
z=e^{\frac{\mu}{T}},
\nonumber\\
\mu=h_{\rm sat}-h,
\;\;\;
h_{\rm sat}=-\epsilon_0=12J_1.
\end{eqnarray}
Here, $\Xi(z,{\cal N})$ is the grand canonical partition function of hard squares on an ${\cal N}$-site square lattice and $z$ is the activity of hard squares.
The contribution, given in  Eq.~(\ref{10}), dominates thermodynamics of the Tasaki-square-lattice $S=1/2$ Heisenberg antiferromagnet at low temperatures around the saturation magnetic field $h_{\rm sat}$.
On the other hand, formula (\ref{10}) allows for calculation of the thermodynamic quantities of the frustrated quantum spin system in this regime by means of the classical statistical mechanics.
In particular, for the specific heat per site of the Tasaki lattice $c(T,h)=C(T,h,N)/N$ and the magnetization per site of the Tasaki lattice $m(T,h)=M(T,h,N)/N$ we easily arrive at the following formulas:
\begin{eqnarray}
\label{11}
C(T,h,N)=z\ln^2\!z\,\frac{\partial\ln\Xi}{\partial z}+z^2\ln^2\!z\,\frac{\partial^2\ln\Xi}{\partial z^2}
\nonumber\\
=\ln^2\!z\,\left(\langle n^2\rangle -\langle n\rangle^2\right),
\nonumber\\
M(T,h,N)=\frac{N}{2}-z\frac{\partial\ln\Xi}{\partial z}
\nonumber\\
=\frac{N}{2}- \langle n\rangle,
\nonumber\\
\langle n\rangle=z\frac{\partial\ln\Xi}{\partial z},
\;\;\;
\langle n^2\rangle=\langle n\rangle{+}\langle n\rangle^2{+}z^2\frac{\partial^2\ln\Xi}{\partial z^2},
\end{eqnarray}
respectively. Here, $\ln z=\mu/T$, $\Xi=\Xi(z,{\cal N})$, and $\langle(\ldots)\rangle=\sum_{n=0}^{{\cal N}/2}z^ng_{\cal N}(n)(\ldots)/\Xi$.
Other thermodynamic quantities, which can be calculated in a similar way, are the entropy $S(T,h,N)=\ln\Xi-z\ln z\,(\partial\ln\Xi/\partial z)$ or the susceptibility $N\chi(T,h,N)=\partial M(T,h,N)/\partial h$, see Ref.~\cite{Derzhko2011}.

\section{Hard squares}
\label{s3}

\subsection{Small finite systems and comparison to exact numerics}
\label{s31}

\begin{figure}
\includegraphics[width=\columnwidth]{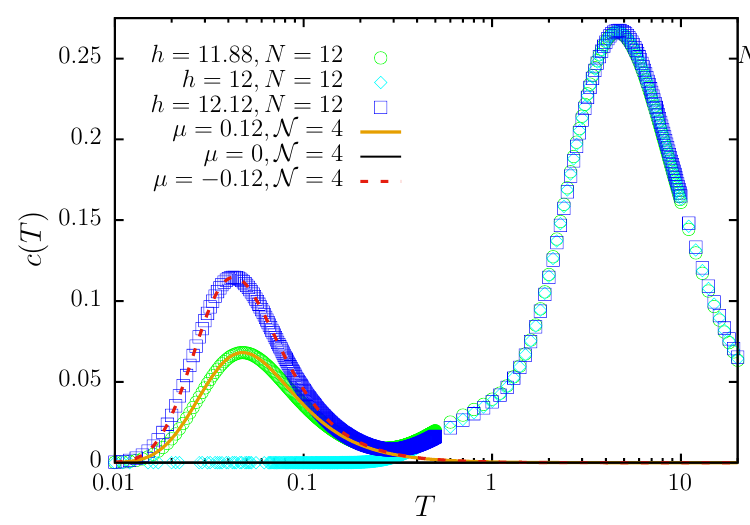} 
\includegraphics[width=\columnwidth]{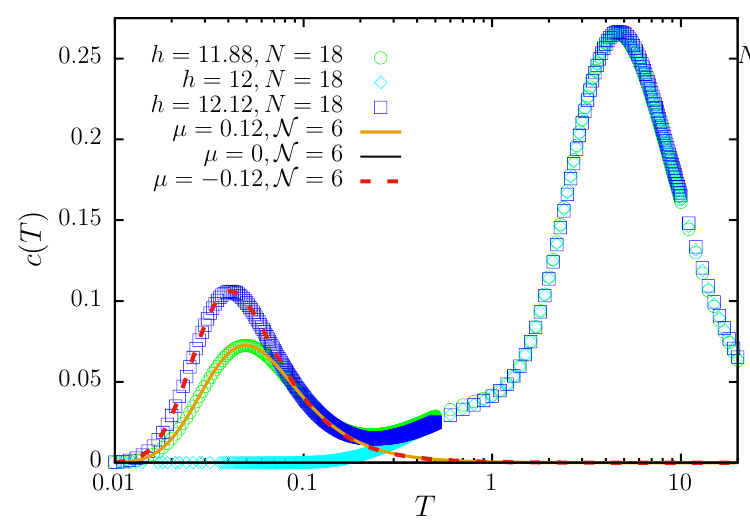} 
\caption{$c(T)$ for the Tasaki square lattice, $J_1=1$, $J_2=4$, $h=11.88,12,12.12$, (top) $N=12$, periodic boundary conditions, and (bottom) $N=18$, twisted/periodic  boundary conditions, see Sec.~\ref{s31}. The results of the localized-magnon description ($\mu=-0.12,0,0.12$ and ${\cal N}=4$ and ${\cal N}=6$) are shown by lines for comparison.}
\label{fig03}
\end{figure}

\begin{figure}
\includegraphics[width=\columnwidth]{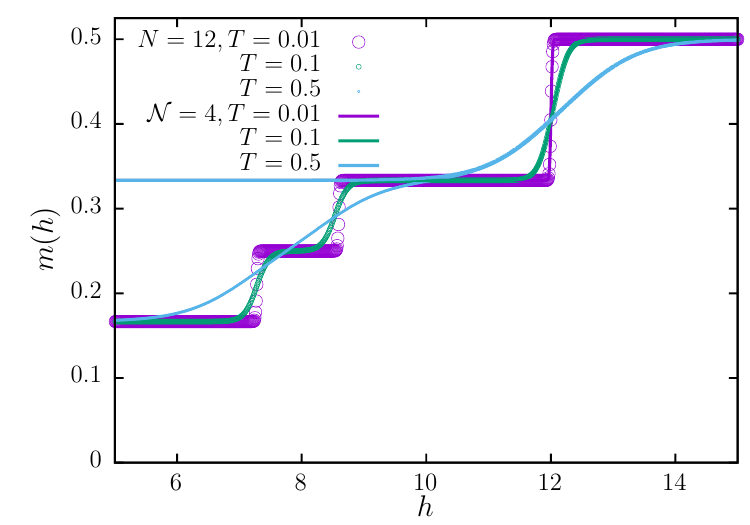} 
\includegraphics[width=\columnwidth]{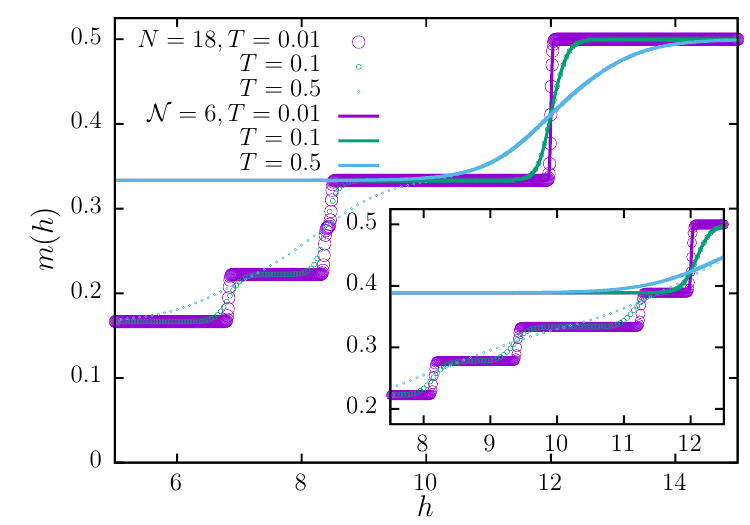} 
\caption{$m(h)$ for the Tasaki square lattice, $J_1=1$, $J_2=4$, $T=0.01,0.1,0.5$,  (top) $N=12$, periodic boundary conditions, and (bottom) $N=18$, twisted/periodic  boundary conditions, see Sec.~\ref{s31}. The results of the localized-magnon description (${\cal N}=4$ and ${\cal N}=6$) are shown by solid lines for comparison. Inset: $N=18$ (${\cal N}=6$), periodic boundary conditions, see Sec.~\ref{s31}.}
\label{fig04}
\end{figure}

To check the localized-magnon description of the low-temperature magnetothermodynamics according to Eqs.~(\ref{10}) and (\ref{11}), we perform exact-diagonalization study \cite{ALBUQUERQUE20071187,Bauer2011} of the small Tasaki clusters of $N=12$ and $N=18$ sites with $J_1=1$ and $J_2=4$. More precisely, $N=3{\cal N}$, ${\cal N}={\cal L}_x{\cal L}_y$, and we consider ${\cal L}_x={\cal L}_y=2$ and ${\cal L}_x=3$, ${\cal L}_y=2$. For ${\cal N}=4$ we impose periodic boundary conditions. For ${\cal N}=6$, besides periodic boundary conditions, which are incompatible with the two-fold degeneracy of the full covering of $6$-site square lattice by $n=3$ hard squares, we impose twisted/periodic boundary conditions in $x$/$y$ direction. 
We compare exact-diagonalization data with data for hard squares on the square-lattice clusters of ${\cal N}=4$ and ${\cal N}=6$ sites.

In Figs.~\ref{fig03} and \ref{fig04} we report the specific heat $c(T)$ around the saturation field $h_{\rm sat}=12$, i.e., at $h=0.99h_{\rm sat}, h_{\rm sat}, h=1.01h_{\rm sat}$, and the magnetization curve $m(h)$ at low temperatures $T=0.01,0.1,0.5$, respectively. Numerics (symbols) are complemented the localized-magnon approach data (lines).
In the latter case we use Eq.~(\ref{10}) with $g_{4}(0)=1$, $g_{4}(1)=4$, $g_{4}(2)=2$ (periodic boundary conditions) and $g_{6}(0)=1$, $g_{6}(1)=6$, $g_{6}(2)=6$, $g_{6}(3)=2$ (twisted boundary conditions). Note that for periodic boundary conditions $g_{6}(0)=1$, $g_{6}(1)=6$, $g_{6}(2)=6$, $g_{6}(3)=0$, i.e., for this specific periodic lattice there is no two full coverings of the ${\cal N}$-site square lattice by $n={\cal N}/2$ hard squares: $g_{6}(3)=0$.
From Figs.~\ref{fig03} and \ref{fig04} we observe a very good agreement between the exact numerics (symbols) and the localized-magnon approach (lines) at low temperatures. For instance, $c(T)$ curves almost coincide until $T$ which is slightly above $T=0.1$, see Fig.~\ref{fig03}. A fingerprint of the localized magnons is a jump in the ground-state magnetization curve at $h=h_{\rm sat}$; the height of the jump is $1/2-1/6=1/3$ (localized magnon states with $n=0,1,2,\ldots,N/6$ magnons have the same energy at $h=h_{\rm sat}$), see the upper panel and the main lower panel of Fig.~\ref{fig04}. 

It is worthwhile noting that for the ${\cal N}=6$ cluster with periodic boundary conditions imposed, both, numerics and hard squares, predict a jump of smaller height, see the inset in Fig.~\ref{fig04}. The obvious reason for this is a peculiarity of the periodic cluster of shape $3\times 2$, for which $g_{6}(3)=0$ instead of expected $g_{{\cal N}}({\cal N}/2)=2$. 
Really, if one considers an infinitely large square lattice without taking care about its boundary, there are precisely two ways to cover it with hard squares: The squares may occupy either the sublattice $A$ or the sublattice $B$ (see the lower panel of Fig.~\ref{fig01}). The periodic cluster with ${\cal L}_x=3$, ${\cal L}_y=2$ does not fit such a pattern. We mention in passing that incompatibility of a finite cluster used in numerics with localized-magnon-crystal pattern may occur for other lattices, too.

Encouraged by this consistency between numerics and effective theory for $N=12$ and $N=18$, see Figs.~\ref{fig03} and \ref{fig04}, we may expect that this agreement preserves for larger $N$ and $N\to\infty$, too. On the other hand, the hard-square model is among well studied models of classical statistical mechanics \cite{Baxter1980,Baxter1980a,Racz1980,Pearce1988,Guo2002}. It can be examined for much larger system sizes using other approaches, like, e.g., classical Monte Carlo simulations. And we may also use this existing knowledge to understand the low-temperature magnetothermodynamics of the initial frustrated quantum spin system. 

\subsection{Large systems and an order-disorder phase transition}
\label{s32}

\begin{figure}
\includegraphics[width=\columnwidth]{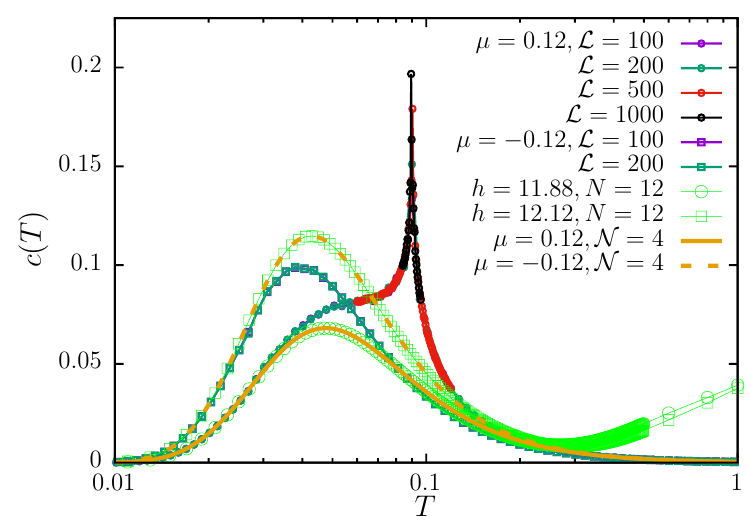} 
\includegraphics[width=\columnwidth]{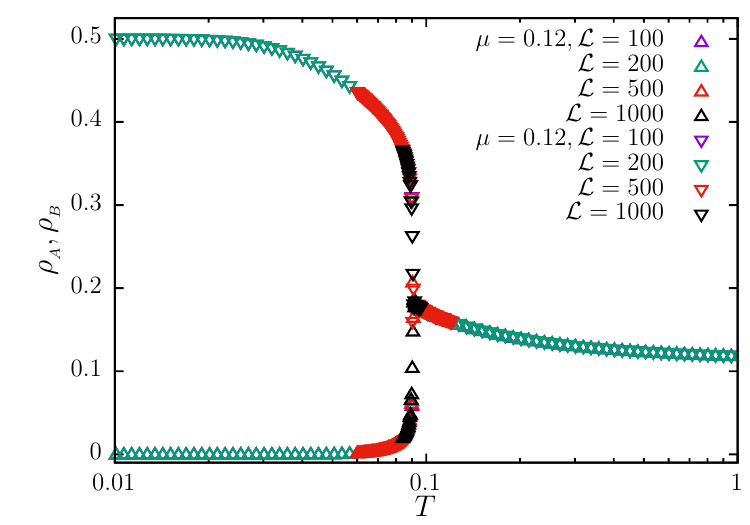} 
\includegraphics[width=\columnwidth]{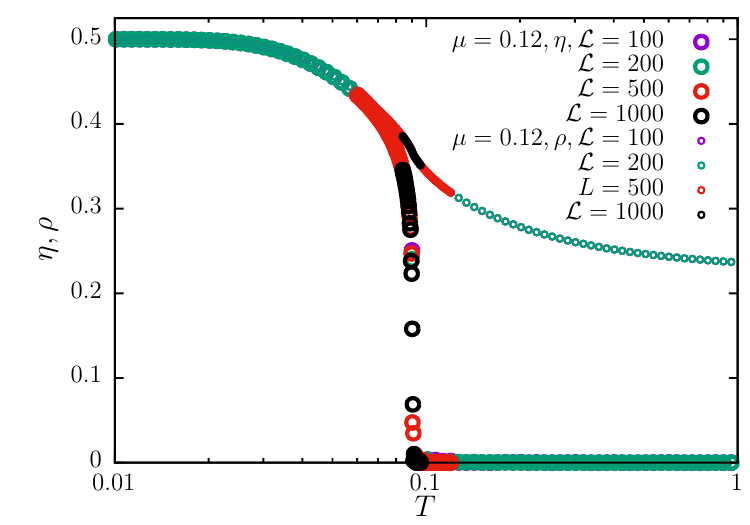} 
\caption{(Top) Expected $c(T)$ for $J_1=1$, $J_2=4$, $h=0.99h_{\rm sat},1.01h_{\rm sat}$, $h_{\rm sat}=12$; classical Monte Carlo simulations for ${\cal N}={\cal L}^2$, ${\cal L}$ up to $1\,000$. 
(Middle) Expected temperature dependence of the sublattice occupation densities $\rho_A=\langle {n}_A\rangle/{\cal N}$ and $\rho_B=\langle {n}_B\rangle/{\cal N}$ at $h=0.99h_{\rm sat}$.
(Bottom) Expected temperature dependence of the order parameter $\eta=\vert \rho_{A}-\rho_{B}\vert$ and the total occupation density $\rho= \rho_{A}+\rho_{B}$ at $h=0.99h_{\rm sat}$.}
\label{fig05}
\end{figure}

In this section, we consider large systems in the same regime of low temperatures around the saturation field. First, we use exact calculations for hard squares up to ${\cal N}=8{\times}8=64$ sites ($N=192$), see Appendix~B. Second, we perform classical Monte Carlo simulations. More precisely, the system sizes were taken up to ${\cal N}=1\,000{\times}1\,000=1\,000\,000$ sites ($N=3\,000\,000$), we used the Metropolis Monte Carlo algorithm, before averaging over $1\,000\,000$ steps, we performed $100\,000$ steps for the initial relaxation, we calculated the averaged values of the numbers of hard squares on the sublattice $A$ ($n_A$) and sublattice $B$ ($n_B$), see the lower panel of Fig.~\ref{fig01}, as well as $(n_A+n_B)^2$, see Eq.~(\ref{11}). Third, we use a broad knowledge about hard squares \cite{Baxter1980,Baxter1980a,Racz1980,Pearce1988,Guo2002}. In particular, it is known that the thermodynamically large hard-square system exhibits an order-disorder phase transition.  

In the top panel of Fig.~\ref{fig05} we report the expected temperature dependence  $c(T)$ for the Tasaki-square-lattice $S=1/2$ Heisenberg antiferromagnet ($J_1=1$, $J_2=4$, $h=0.99h_{\rm sat}$, $h=1.01h_{\rm sat}$, $h_{\rm sat}=12$) as it follows from classical Monte Carlo simulations for hard squares ($\mu=0.12$ and $\mu=-0.12$). Visibly, $c(T)$ exhibits a peculiarity at $T\approx 0.09$. 
In the middle and bottom panels of Fig.~\ref{fig05} we report temperature dependences of the sublattice occupation densities $\rho_A=\langle {n}_A\rangle/{\cal N}$ and $\rho_B=\langle {n}_B\rangle/{\cal N}$ (triangles) along with the order parameter $\eta=\vert \rho_A-\rho_B\vert$ (large circles) and the total occupation density  $\rho= \rho_A+\rho_B$ (small circles). The presented temperature dependencies are for fixed $\mu=0.12$, i.e., $h=0.99h_{\rm sat}$.
At high temperatures for some fixed $\mu>0$, the activity $z=e^{\mu/T}$ is only slightly above 1 and both sublattices are equally occupied. With temperature decrease, the activity $z$ grows and just above the critical value $z_c=3.796\,2\pm 0.000\,1$ \cite{Baxter1980a}, i.e., when $T$ is $\mu/\ln z_c\approx 0.12/1.334\approx0.09$, one of two sublattices becomes occupied more than the other one, compare up- and down-triangles in the middle panel of Fig.~\ref{fig05}. As a result, the order parameter $\eta$ (large circles in the lower panel of Fig.~\ref{fig05}) becomes nonzero signalizing the entering to the ordered phase. 
Since a hard square corresponds to a localized magnon occupying five sites of the Tasaki square lattice and obeying hard-square exclusion rule, the order-disorder phase transition corresponds to the ordering of the independent localized magnons at low temperature just below the saturation magnetic field: As their density increases they (spontaneously) start to occupy one of two sublattices of the background square lattice, see Fig.~\ref{fig01}. 
Finally, we emphasize again that 
the critical temperature for the discussed order-disorder phase transition reads:
\begin{eqnarray}
\label{12}
\left. T_c(h)\right\vert_{h\to h_{\rm sat}-0}=\frac{h_{\rm sat}-h}{\ln z_c},
\;\;\;
\ln z_c \approx 1.334
\end{eqnarray}
and it is about $0.09$ for $h=11.88$ for the chosen set of parameters $J_1=1$ and $J_2=4$ when $h_{\rm sat}=12$.
Moreover, the phase transition in question belongs to the 2D Ising universality class \cite{Racz1980}.

\section{Discussion and outlook}
\label{s4}

In the present paper, we have examined the low-temperature magnetothermodynamics of a frustrated quantum Heisenberg antiferromagnet, which supports a dispersionless (flat) magnon band, namely, the Tasaki-square-lattice $S=1/2$ Heisenberg antiferromagnet. The idea of our approach is to map low-energy states onto spatial configurations of hard-core objects on an auxiliary lattice and then to use the classical statistical mechanics to find the low-temperature magnetothermodynamics.  
Comparison for small systems gives evidence that the mapping works, and, therefore, an order-disorder phase transition is expected in frustrated quantum spin system in accordance with Eq.~(\ref{12}). We have to mention here that such a way to calculate thermodynamic quantities is not new \cite{Zhitomirsky2004,Derzhko2006} and has been realized several times during the last two decades.
With our study, we have added to a list of systems belonging to the hard square universality class \cite{Richter2006,Derzhko2010,Richter2018} one more frustrated quantum spin system -- the Tasaki-square-lattice $S=1/2$ Heisenberg antiferromagnet.

Let us mentioned a few issues which, in our opinion, deserve further discussion.
As has been mentioned in  Sec.~\ref{s2}, the model exhibits one more flat band (see Fig.~\ref{fig02}), which in the case of Tasaki flat-band point $J_2=4J_1$ is not the lowest-energy one and therefore is irrelevant for the ground states in the subspaces with total $S^z=N/2,\ldots,N/3$. However, the flat-band states with the energy $\Lambda_{\bm q}^{(2)}=-4J_1$ may be relevant for higher temperatures/lower magnetic fields. Understanding the role of the other flat band seems to be an interesting issue for future study.
Another topic for further study may be a similar analysis of other two-dimensional Tasaki lattices, which emerge after, say, the honeycomb or triangular lattice is  decorated in the style  of Tasaki. 

Finally, it is worthy comparing two different in nature flat-band systems, the antiferromagnetic Heisenberg model and the standard Hubbard model, both on the Tasaki square lattice. On the one-particle level they share many similarities: If the hopping integrals $t_2=\sqrt{z}t_1>0$ ($t_1$ is the hopping integral along the square lattice, $t_2$ is the hopping integral along decoration, $z=4$ is the coordination number of the underlying square lattice), there is a flat band and the one-electron states in the flat band can be taken as localized on traps shown in Fig.~\ref{fig01}. The flat-band energy is $-zt_1=-4t_1$. However, a localized eigenstate, besides the index corresponding to the site of the underlying square lattice, has the spin index $\sigma=\uparrow,\downarrow$. Many-electron ground states are not only those which are visualized as electrons in distant traps with arbitrary spin indices, but also those which are visualized as electrons in traps with common site(s), however, under requirement of correlated spin index -- only in such a case the Hubbard repulsion $U>0$ does not increase their energy.
Furthermore, many-particle ground states can be mapped onto a hard-square problem (Heisenberg case) or a percolation problem (Hubbard case). In both cases the {\it classical} statistical mechanics tools are applicable for examine quantum lattice system in the specific regime of high magnetic field (around the saturation) or
low electron density (half-filled flat band).

Our theoretical consideration may be also of interest from the solid-state physics perspective. Unfortunately, we are not aware about any realization of the (spin or electron) Tasaki lattice in real-life compounds but we should not dismiss the possibility of such compounds being discovered/synthesized at some future time. Cold atoms in optical lattices may be another setup for experimental observation of flat-band physics in the Tasaki lattice.

\section*{Data availability statement}

The data that support the findings of this study are available from the authors upon reasonable request.

\section*{Acknowledgements}

The authors are thankful to the Armed Forces of Ukraine for protection since 2014, and especially since February 24, 2022. 
This project is funded by the National Research Foundation of Ukraine (2023.03/0063, Frustrated quantum magnets under various external conditions); the authors are indebted to T.~Hutak and T.~Verkholyak for useful discussions. 
O.~D. thanks the Abdus Salam International Centre for Theoretical Physics (Trieste) for kind hospitality at the Joint ICTP-WE Heraeus School and Workshop on Advances in Quantum Matter: Pushing the Boundaries, 4 -- 15 August 2025.
O.~D. also thanks A.~Honecker, J.~Stre\v{c}ka, and K.~Karľov\'{a} for kind hospitality at the COOLMAG2025 Workshop Magnetic Cooling and Frustrated Magnetism, October 27th -- 29th, 2025, Ko\v{s}ice, Slovakia.

\section*{Appendix~A: One-magnon states}
\renewcommand{\theequation}{A.\arabic{equation}}
\setcounter{equation}{0}

In Sec.~\ref{s2} we report the eigenstates $\vert\psi^{(3)}_{\bm q}\rangle$, see Eq.~(\ref{05}). Two other eigenstates are as follows:
\begin{eqnarray}
\label{A1}
\vert\psi^{(1)}_{\bm q}\rangle
{=}
\left[\frac{2{+}\cos q_x{+}\cos q_y}{2(1{+}e^{{\rm i}q_y})}S^-_{{\bm q}1}
{+}\frac{1{+}e^{{\rm i}q_x}}{1{+}e^{{\rm i}q_y}}S^-_{{\bm q}2}
{+}S^-_{{\bm q}3}
\right]\vert{\rm FM}\rangle
\end{eqnarray}
and
\begin{eqnarray}
\label{A2}
\vert\psi^{(2)}_{\bm q}\rangle
=
\left[-\frac{1+e^{-{\rm i}q_y}}{1+e^{-{\rm i}q_x}}S^-_{{\bm q}2}+S^-_{{\bm q}3}\right]\vert{\rm FM}\rangle;
\end{eqnarray}
they correspond to the energies $\Lambda_{\bm q}^{(1)}$ and $\Lambda_{\bm q}^{(2)}$, see Eq.~(\ref{04}), respectively. We present these eigenstates here for completeness. 

Note that Eq.~(\ref{A2}) with the help of Eq.~(\ref{02}) can be cast into
\begin{eqnarray}
\label{A3}
\vert\psi^{(2)}_{\bm q}\rangle
=
\frac{1}{\sqrt{{\cal N}}}\sum_{{\bm m}}e^{{\rm i}{\bm q}{\cdot}{\bm m}}\frac{1}{1+e^{-{\rm i}q_x}}L_{\bm m}\vert {\rm FM}\rangle,
\nonumber\\
L_{\bm m}={-}S_{{\bm m}2}^-{+}S_{{\bm m}3}^-{-}S_{m_x-1,m_y,2}^-{+}S_{m_x,m_y-1,3}^-,
\end{eqnarray}
cf. Eqs.~(\ref{05}) and (\ref{06}),
and because $\Lambda_{\bm q}^{(2)}=-4J_1$ does not depend on ${\bm q}$, these flat-band states can be written as located on four sites only, i.e., $L_{\bm m}\vert {\rm FM}\rangle$, ${\bm m}$ runs over ${\cal N}$ sites of the underlying square lattice. We do not discuss further the localized-magnon states given in Eq.~(\ref{A3}) and leave this task for future studies. 
In contrast, the states from the dispersive band $\vert\psi^{(1)}_{\bm q}\rangle$  given in Eq.~(\ref{A1}), cannot be taken as the ones located on a small part of the lattice since their energies $\Lambda_{\bm q}^{(1)}$ (\ref{04}) do depend on ${\bm q}$.

\section*{Appendix~B: Hard squares on square lattice}
\renewcommand{\theequation}{B.\arabic{equation}}
\setcounter{equation}{0}

\begin{figure}[b]
	\includegraphics[width=\columnwidth]{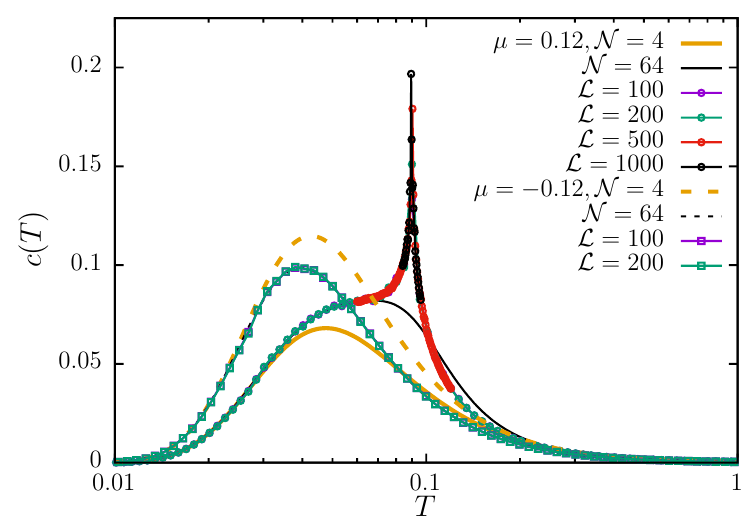}
	\caption{Expected $c(T)$ for $J_1=1$, $J_2=4$ ($h_{\rm sat}=12$), $h=0.99h_{\rm sat}$ ($\mu=0.12$) and $h=1.01h_{\rm sat}$ ($\mu=-0.12$); direct calculations for ${\cal N}=4$, ${\cal N}=64$, and classical Monte Carlo simulations for up to ${\cal N}=1\,000\,000$, cf. top panel of Fig.~\ref{fig05}.}
	\label{fig06}
\end{figure}

In this appendix we explain how to calculate the canonical partition function of $n=1,\ldots,{\cal N}/2$ hard squares on a square lattice of ${\cal N}=\sqrt{{\cal N}}{\times}\sqrt{{\cal N}}$ sites, denoted as $g_{\cal N}(n)$; here ${\cal N}$ is assumed to be even. To this end, we vary the occupation of each lattice site $j=1,\ldots,{\cal N}$, $n_j=0,1$, and check whether a configuration $\{n_1,\ldots,n_{\cal N}\}$ satisfies the hard-core rules (we are prepared to inspect $2^{\cal N}$ configurations, but the actual number of verification is much smaller). For each allowed configuration we remember the number of occupied sites $n$. At the end of the day, we get the required functions $g_{\cal N}(n)$ for $n=0,1,\ldots,{\cal N}/2$.

We illustrate such calculations by considering the case of $64=8\times 8$ sites, periodic boundary conditions are imposed, for which the grand canonical partition function $\Xi(z,64)=\sum_{n=0}^{32}g_{64}(n)z^n$ is as follows:
\begin{eqnarray}
\label{b1}
\Xi(z,64)= 1{+}64z{+}1\,888z^2{+}34\,112z^3{+}423\,152z^4
\nonumber\\
{+}3\,830\,016z^5{+}26\,249\,184z^6{+}139\,580\,160z^7
\nonumber\\
{+}585\,632\,520z^8
{+}1\,962\,132\,800z^9{+}5\,296\,005\,568z^{10}
\nonumber\\
{+}11\,591\,943\,552z^{11}{+}20\,681\,906\,352z^{12}
\nonumber\\
{+}30\,206\,108\,416z^{13}{+}36\,251\,041\,536z^{14}
\nonumber\\
{+}35\,886\,874\,048z^{15}{+}29\,436\,488\,660z^{16}
\nonumber\\
{+}20\,127\,048\,512z^{17}{+}11\,573\,937\,440z^{18}
\nonumber\\
{+}5\,674\,532\,608z^{19}{+}2\,420\,605\,568z^{20}
\nonumber\\
{+}922\,331\,136z^{21}{+}322\,239\,232z^{22}
\nonumber\\
{+}104\,747\,904z^{23}{+}31\,534\,744z^{24}
\nonumber\\
{+}8\,617\,024z^{25}{+}2\,080\,576z^{26}{+}430\,848z^{27}
\nonumber\\
{+}73\,840z^{28}
{+}9\,984z^{29}{+}992z^{30}{+}64z^{31}{+}2z^{32}.	
\end{eqnarray}
Knowing $\Xi(z,64)$ (\ref{b1}), one immediately gets the hard-square predictions for the Tasaki-square-lattice $S=1/2$ Heisenberg antiferromagnet of $N=192$ sites around the saturation field, see Eqs.~(\ref{10}) and (\ref{11}). 

In Fig.~\ref{fig06} we illustrate $c(T)$ at $\mu=\pm0.12$ as it follows from $\Xi(z,64)$ (\ref{b1}). Clearly, the 64-site system cannot illustrate a logarithmic singularity to develop at $T_c=0.12/\ln z_c$ below the saturation field (when $\mu$ is positive), but comparing the data for ${\cal N}=4$, ${\cal N}=64$ to Monte Carlo simulations for ${\cal N}=1\,000\,000$ the correct tendency as ${\cal N}$ grows is obvious. In contrast, above the saturation field (when $\mu$ is negative), there are almost no difference between ${\cal N}=64$ and ${\cal N}=40\,000$ data. 
               
\bibliography{tasaki_refs}

\end{document}